\documentclass{revtex4}
\usepackage{graphicx}
\usepackage{fancyhdr}
\usepackage{amsmath}
\usepackage{color}

\begin{document}

\title{Thermal Phase Transition in the $SO(5)\times U(1)$ Gauge-Higgs Unification with 126GeV Higgs}


\author{H. Hatanaka}
\affiliation{Department of Physics, Osaka Uniersity, Toyonaka, Osaka, JAPAN}

\begin{abstract}
We study the phase structure of the gauge theories in the space-time with one compact dimension, where the gauge symmetry can be broken by the Hosotani mechanism.
As the extra dimension, we consider the $SO(5)\times U(1)$ gauge-Higgs unification in the Randall-Sundrum space-time which reproduce the $126$ GeV Higgs mass.
It is found that the thermal phase transition of the electroweak symmetry is very weak first order or almost second order and that the critical temperature is around $160$ GeV for $z_L \lesssim 10^7$ and $n_F=3$.
(This work is in progress.)
\end{abstract}

\maketitle

\thispagestyle{fancy}


%%%%%%%%%%%%%%%%%%%%%%%%%%%%%%%%%%
\section{Introduction}

Higgs physics may be related the origin of the baryon asymmetry of the universe.
In the electroweak baryogenesis scenario\cite{Sakharov:1967dj},
 in order to generate the baryon number in the electroweak phase transition,
it is necessary that the phase transition is strong first-order.
This criterion is quantitatively described as \cite{Klinkhamer:1984di}
\begin{eqnarray}
\frac{\varphi_c}{T_c} \gtrsim 1,
\label{shapo}
\end{eqnarray}
where $T_c$ is the critical temperature of the electroweak phase transition
and $\varphi_c$ is the magnitude of the 
vacuum expectation value (VEV) of the Higgs at $T=T_c$. 

It is known that in the standard model (SM) the phase transition is second-order 
for $m_H \gtrsim 70$ GeV.
Thus if the Higgs mass is $126$ GeV as overfed in LHC,
it is difficult to obtain the baryon asymmetry through the electroweak phase transition (EWPT).
Therefore, in order to make the electroweak baryogenesis work successfully,
the Higgs sector in the SM should be extended.

In this presentation we briefly report the EWPT in a particular model of the gauge-Higgs unification (GHU)\cite{Funatsu:2013ni}.
In the gauge-Higgs unification scenario\cite{GHU},
the Higgs field is regarded as the zero-mode of the extra-dimensional component of the gauge field in the higher-dimensional space-time.
The gauge symmetry breaking is triggered by the non-trivial vacuum expectation value (VEV) of the
Wilson-line phase. This is what we call the Hosotani mechanism\cite{Hosotani:1983xw}.
It has been reported that in GHU models the thermal phase transition is strong first order\cite{GHU-FT}. 
In this presentation we report the EWPT in a realistic GHU model which reproduce
appropriate Higgs mass $126$ GeV.

\section{Thermal phase transition in the $SO(5)\times U(1)$ gauge-Higgs unification with $m_H=126\text{ GeV}$ Higgs}

We consider the model in \cite{Funatsu:2013ni}. This is 
an $SO(5)\times U(1)_X$ gauge theory in the Randall-Sundrum space-time.\cite{Randall:1999ee}
\begin{eqnarray}
ds^2 = e^{-2k|y|}(\eta_{\mu\nu} dx^\mu dx^\nu) - dy^2,
\quad
-L \le y \le L,
\end{eqnarray}
where $k$ is the curvature of the five-dimensional anti-de Sitter space.
The $SO(5) \times U(1)_X$ gauge symmetry is broken to $SU(2) \times U(1)_Y$
by the boundary conditions at $y=0,\pm\pi R$.
The electroweak symmetry breaking is  by the VEV of the 5-dimensional component of the gauge field $A_y$ in the direction of $SO(5)/SO(4)$. The magnitude of the VEV is parameterized by the Wilson-line phase $\theta_H$ \cite{Hosotani:2008tx}: 
$\exp[\frac{i}{2}\theta_H (2\sqrt{2}T^{\hat{4}})] = \exp[i g_A] \int_1^{z_L} dz \langle A_z
\rangle$ ($z = e^{ky}$, $z_L \equiv e^{kL}$).

\begin{figure}[htbp]
\centerline{\includegraphics[width=8cm]{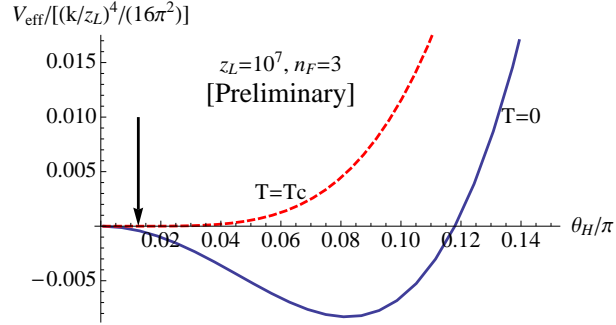}}
\caption{The effective potential with respect to the Wilson-line phase $\theta_H$ for $z_L=10^7$ and $n_F=3$, in the unit of $(k/z_L)^4/(16\pi^2)$ where $k=1.26\times10^{10}$ GeV \cite{Funatsu:2013ni}. The blue-solid [red-dashed] line is for $T = 0$ [$T=T_c=163$ GeV].
The downward-arrow indicates the position of the nontrivial minimum at $T=T_c$. }\label{fig-veff07}
\end{figure}

In order to study the phase structure of this model,
we evaluate the effective potential at finite temperature.
The effective potential consists of zero-tempereture effective potential and 
the finite-temperature corrections:
\begin{eqnarray}
V_{\rm eff} &=& V_{\rm eff}^{T=0} + \Delta V_{\rm eff}.
\end{eqnarray}
The zero temperature part of the effective potential $V_{\rm eft}^{T=0}$
is calculated by the method developed in \cite{effpot}.
The explicit forms of $V_{\rm eft}^{T=0}$ in this model can be seen in Refs.~
\cite{Hosotani:2008tx,Funatsu:2013ni}.
Finite temperature corrections are given by
\begin{eqnarray}
\Delta V_{\rm eff}
&=& 
- \frac{T^4}{2\pi^2} \biggl\{
G[ \{ m_{n}^{(W)} \} ,0] + G[ \{ m_{n}^{(Z)} \} ,0] + G[ \{ m_{n}^{(H)} \},0]
\nonumber\\&& 
+ G[ \{ m_{n}^{(t)}\},\tfrac{1}{2}] 
+ G[ \{ m_{n}^{(F)}\}, \tfrac{1}{2}] 
\biggr\},
\\
G[\{ M_n \},\eta] &=& (-1)^{2\eta} \sum_{m=1}^{\infty} \sum_{n} \frac{(-1)^{2m\eta}}{m^d}
B_2 (m M_n / T),
\quad B_2(x) \equiv x^2 K_2(x),
\end{eqnarray}
where the KK mass spectra $\{m_n^{X}\}$ ($X=W,Z,H,t,F$) is for the W-tower, Z-tower, the scalar-tower, top-tower and the KK-tower of the $SO(5)$-spinorial fermions, respectively, and They are given in \cite{Hosotani:2008tx,Funatsu:2013ni}.
This model contains two free parameters, namely, the warp factor $z_L$ and the number of the spinorial-representation fermion $n_F$. In this report we consider the case with $n_F=3$. The effective potential is numerically evaluated.
In Fig.~\ref{fig-veff07}, the $V_{\rm eff}$ with respect to $\theta_H$
at $T=0$ and $T=T_C=163$ GeV for $(z_L, n_F) = (10^7,3)$ are plotted.
We note that at $T=T_C$ there are two minima of the potential, one of which is $\theta_H=0$ and the other is $\theta_H \sim 0.012\pi$.

\begin{figure}[htbp]
\centerline{\includegraphics[width=8cm]{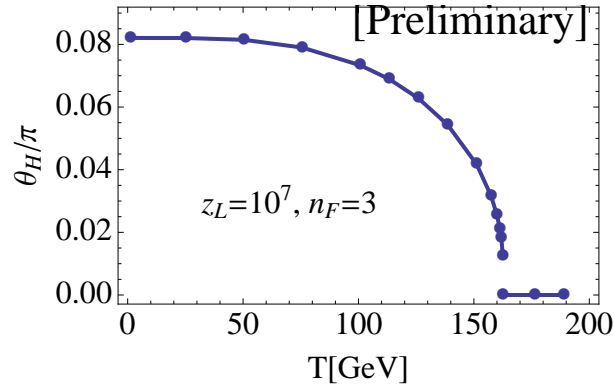}}
\caption{%
Position of the minimum of the effective potential v.s. temperature $T$ for $z_L = 10^7$ and $n_F=3$. At $T \sim 163$ GeV a tiny gap is observed.}\label{fig-ph07}
\end{figure}

In Fig. \ref{fig-ph07}, the position of the minimum of the effective potential, $\theta_H$ with respect to the temperature $T$ for $(z_L,n_F)=(10^7,3)$ is plotted.
The position $\theta_H$ monotonically decreases as $T$ increases, and there are
tiny gap at $T=163\text{ GeV}=T_C$.
When $T> T_C$, $\theta_H$ vanishes and the $SU(2)\times U(1)_Y$ symmetry restores.
This is a first order phase transition but it also seems very weak and almost second order.

In order to check the criterion (\ref{shapo}),
we define $\varphi_c$ as
\begin{eqnarray}
\varphi_c \equiv \frac{\theta_H(T=T_c)}{\theta_H(T=0)} \cdot 246\text{ GeV}.
\end{eqnarray}
where $\theta_H(T=T_c)$ is the position of the non-trivial minimum of $V_{\rm eff}$ at $T=T_C$.
In Table \ref{tab-td}, we summarized $T_c$, $\varphi_c$ for the various value of $z_L$ for
$n_F=3$.
\begin{table}[htbp]
\caption{Critical temperatures, magnitudes of the vacuum expectation value $\varphi_c$
for various values of $z_L$ with $n_F=3$.}\label{tab-td}
\begin{center}
\begin{tabular}{lll}
\hline
$z_L$ & $T_c$ [GeV] & $\varphi_c$ [GeV] \\
\hline
$2\times10^4$ & $168$ & $12.8$ \\
$10^{5}$ & $161$ & $12.2$\\
$10^{7}$ & $163$ & $11.9$\\
$10^{10}$ & $190$ & $16.3$\\
$10^{12}$ & $215$ & $26.3$\\
\hline
\end{tabular}
\end{center}
\end{table}
$T_c$ varies from $160$ GeV to $215$ GeV for $2\times 10^4 \le z_L \le 10^{12}$.
In particular $T_C \sim {\cal O}(160\text{GeV})$ for $z_L \lesssim 10^7$.
The values $\varphi_C/T_C \sim {\cal O}(0.1)$ are much smaller than unity for all values of $z_L$.
Therefore this model does not fulfill the criterion Eq. (\ref{shapo}) for $n_F=3$.

\section{Summary and Discussion}
The thermal phase transition of the $SO(5)\times U(1)$ gauge-Higgs unification model which reproduce the $126$ GeV Higgs mass was studied.
The effective potential of the gauge-Higgs unification at finite temperature was 
formulated, and phase structure of the $SO(5) \times U(1)$ gauge-Higgs unification model was studied.
It is found that the thermal phase transition in the model is very weak first order or almost second-order. 
This implies the electroweak baryogenesis cannot occur successfully in this model because the baryon-number is washed-out by the sphaleron.

The result does not exclude the possibility to reproduce the baryon number by other mechanism,
e.g. leptogenesis. It might be possible to make the first-order phase transition stronger
by introducing supersymmetry into the model\cite{Hatanaka:2011ru}, in analog with
the enhanced first-order phase transition in the SM with supersymmetry\cite{MSSM}.

In this presentation an analysis only for the $n_F=3$ case is reported. Analysis including result for other values of $n_F$ will be summarised in somewhere else.

\begin{acknowledgments}
This work was supported in part by Grant-in-Aid for Scientific Research (A) 20244028
from the Ministry of Education and Science (MEXT) of Japan.
\end{acknowledgments}

\bigskip % extra skip inserted

\end{document}